
\def\gsim{\mathrel{\scriptstyle{\buildrel > \over \sim}}}

\magnification 1200
\baselineskip=17pt

\centerline{\bf THERMAL PROPERTIES OF GAUGE-FIELDS COMMON}
\bigskip
\centerline{\bf TO ANYON SUPERCONDUCTORS AND SPIN-LIQUIDS}
\vskip 50pt
\centerline{J. P. Rodriguez}
\medskip
\centerline{\it Theoretical Division,
Los Alamos National Laboratory,
Los Alamos, NM 87545 and}
\centerline{{\it Dept. of Physics and Astronomy,
California State University,
Los Angeles, CA 90032.}\footnote*{Permanent address.}}
\vskip 30pt
\centerline  {\bf  Abstract}
\vskip 8pt\noindent
The thermally driven confinement-deconfinement 
transition exhibited by
lattice quantum electrodynamics in two space dimensions
is re-examined in the context of the statistical
gauge-fields common to  anyon superconductors 
and to spin-liquids.   Particle-hole excitations
in both systems are bound by a confining string
at temperatures below  the transition temperature $T_c$.
We argue that $T_c$ coincides with 
the actual critical temperature for anyon
superconductivity.  The corresponding
specific-heat contribution, however,  
shows a {\it smooth} peak just below
$T_c$ characteristic of certain high-temperature superconductors.

\bigskip
\noindent
PACS Indices:  74.20.Kk, 74.20.Mn, 11.15.Ha, 71.27.+a
\vfill\eject
The discovery of high-temperature superconductivity$^1$ 
has generated a number of new theoretical approaches 
to the problem of strongly interacting Fermi systems 
in two dimensions.  Among these, the proposals of spin-liquids in two-dimensional (2D) Mott insulators and of anyon superconductivity
in doped 2D Mott insulators are perhaps the most novel.$^{2-4}$
The former state is characterized by short-range spin correlations
while the latter is essentially a 2D superconductor that generally
breaks time-reversal and parity symmetries.$^4$  In addition,
both the spin-liquid state and the anyon superconductor have
a gap for Fermi excitations.$^4$

Parallel technical developments  
have evolved in the arena of $U(1)$ gauge-field
descriptions of strongly correlated electron systems 
in two dimensions.  In particular, the 2D spin-1/2
antiferromagnet has been treated within both the
Schwinger-boson scheme$^5$ and the corresponding
long-wavelength description given by the $CP_1$ model,$^{6,7}$
where the  gauge-field that appears
in such case measures chiral spin-fluctuations.$^8$
In addition, the random-phase approximation (RPA) results
for anyon superconductors$^9$ have been recovered by Chern-Simons
(CS) gauge-theories in the mean-field approximation,$^{10}$ where
a statistical gauge-field is introduced to describe the 
flux-tube attached to each fermion by the CS
term.   Very similar gauge-field descriptions
also exist for anyon superconductors 
in the context of doped 2D antiferromagnets
($t-J$ model).$^{11,12}$
In both the case of the spin-liquid (quantum disordered)
phase of the 2D spin-1/2 antiferromagnet and of the
anyon superconductor, fluctuations in the (statistical) gauge-field
are described by an action for {\it vacuum} 
quantum electrodynamics
in 2+1 dimensions (QED$_3$).$^{6,7,12}$  
This does not occur in
the case of either the 2D  N\'eel phase$^7$ or of the
``strange'' metal phase of the 2D $t-J$ model,$^{13}$
where spin excitations
are gapless.

In this paper,  we study the thermal properties of such
statistical gauge-fields common to anyon superconductors and to
spin-liquids on the square lattice.  Specifically, we re-examine
the confinement-deconfinement (CD) transition
at non-zero temperature
experienced by particle-hole excitations$^{14,15}$ about
such groundstates using an effective 
lattice (compact) QED$_3$
lagrangian for the statistical gauge-field.   
The CD transition is found to be
dual to the 2D Coulomb-gas (CG) transition
in the weak-coupling regime, 
like in the strong-coupling regime.$^{16,17}$
In fact, both the string-tension and the 
inverse of the confinement length-scale vanish exponentially as 
the transition temperature, $T_c$,
is approached from low-temperature at both the
strong and weak-coupling limits.  The specific-heat
is also computed,
where it is characterized by a {\it smooth} bump
anomaly below $T_c$ reminiscent of certain high-temperature
superconductors.$^{18}$
Last, we argue that $T_c$ coincides with the 
actual transition
temperature for anyon superconductivity.

The action for compact QED$_3$ 
at temperature $T\neq 0$ is defined by
$$S={1\over{2 g_0^2}}\sum_{x_3/a=1}^{\beta}\sum_{\vec x}
\sum_{\mu,\nu=1,2,3}\{1-{\rm cos}[\Delta_{\mu}a_{\nu}(x)
-\Delta_{\nu}a_{\mu}(x)]\}, \eqno (1)$$
where $a_{\nu}(x)$ represents the statistical gauge-field,
$\Delta_{\mu}$ denotes the lattice difference operator,
and where $x=(x_1,x_2,x_3)$ is the three-vector that spans
the cubic-lattice space-time with lattice constant $a$.
Quantum statistical mechanics requires that the gauge-field
be periodic in the time-like direction$^{19}$ $x_3= i c_0 t$,
where $c_0$ is the zero-sound speed for anyon superconductors
and the spin-wave velocity for spin-liquids; i.e., 
$a_{\nu}(x_3+\beta a)=a_{\nu}(x_3)$ mod $2\pi$, where 
$\beta=\hbar\omega_0/k_B T$ is presumed to be a positive integer,
with $\omega_0= c_0 a^{-1}$ as the Debye frequency-scale
of the theory.  Polyakov has shown that compact QED$_3$
is confining at zero-temperature in the weak-coupling
regime,$^{20,21}$ $g_0\ll 1$.  Below, we extend his instanton-gas
analysis to the present case of non-zero temperature, and find that 
a CD-transition occurs at 
$k_B T_c \sim  g_0 \hbar\omega_0$.

{\it Weak-coupling.}  In the limit  $g_0\ll 1$, 
the Villain substitution
for the partition function, 
$Z=\int {\cal D} a_{\mu} e^{-S}$, is then valid,$^{20-22}$
yielding the factorization $Z=Z_{\rm gauss}Z_{\rm inst}$,
where $Z_{\rm gauss}$ represents the gaussian (``spin-wave'')
approximation to the action (1),
and where the corresponding instanton contribution
is
$$Z_{\rm inst} = \sum_{\{n(\bar x)\}}{\rm exp}
\Biggl\{-{g_0^2\over 2}\sum_{\bar x}^{\qquad\prime}
[\Delta_{\mu} n(\bar x)]^2\Biggr\}.\eqno (2)$$
Above, $n(\bar x)$ is an integer-field on the dual cubic-lattice,
such that
$n_{\mu\nu}(x)=\epsilon_{\mu\nu\lambda}\Delta_{\lambda} n(\bar x)$
is the  antisymmetric integer-field dual to the minima of the
periodic energy functional (1) at a given plaquette $(\mu,\nu)$.
[The link associated with $\Delta_{\lambda} n(\bar x)$ passes
perpendicularly through the plaquette associated  with 
$n_{\mu\nu}(x)$.]   Again, the field
$n(\bar x)$ must be periodic in the time-like direction 
with period $\beta a$; i.e.,  the sum 
over time $\bar x_3$ in Eq.  (2)
is restrict to a slab of thickness $\beta a$.
After employing the Poisson summation formula 
on the configuration sum above (2), and then 
integrating over the  continuous field corresponding
to $n(\bar x)$, we arrive at the following instanton-gas 
ensemble:$^{20-22}$
$$Z_{\rm inst} = \sum_{\{m(\bar x)\}}{\rm exp}
\Biggl[-{1\over{2 g_0^2}}(2\pi)^2
\sum_{\bar x, \bar x^{\prime}}^{\qquad\prime}
m(\bar x) G^{(3)}(\bar x -\bar x^{\prime}) m(\bar x^{\prime})
\Biggr], \eqno (3)$$
where $G^{(d)}(\bar x)= (2\pi)^{-d}\int_{\rm BZ}d^dk e^{ik\cdot \bar x}
[2d-2\sum_{\mu=1}^d {\rm cos} (k_\mu a)]^{-1}$ is the Greens function
for the $d$-dimensional hypercubic lattice, and where $m(\bar x)$ is the  integer
charge-field of the instanton-gas, which is also periodic
in the time-like direction.  Consider now two instantons of charge
$m(\bar x)$ and $ m(\bar x^{\prime})$ centered at space-time points
$\bar x=(\vec r, \bar x_3)$ and $\bar x^{\prime}=(\vec r\,^{\prime},
\bar x_3\,^{\prime})$, respectively. 
Furthermore, suppose that they are widely spaced, such that
$|\vec r-\vec r\,^{\prime}|\gg \beta a$.
Then since both configurations
have period $\beta a$ along the time-like direction, their mutual
action (3) per elementary time-slice is proportional
to 
$\beta^{-1}m(\vec r) G^{(2)}(\vec r -\vec r\,^{\prime}) \beta^{-1}m(\vec r\,^{\prime})$, where $G^{(2)}(\vec r)$ is the Greens function
for the square-lattice and  $\beta^{-1}m(\vec r)$ is simply the 
linear charge density along time.  After multiplying the
latter action-density by $\beta$ to get the action
per period, we arrive at the following 2D CG partion
function for the periodic instanton gas
ensemble (3):
$$Z_{\rm inst} = \sum_{\{m(\vec r)\}}
\beta^{N_m}{\rm exp}
\Biggl[-{1\over{2\beta g_0^2}}(2\pi)^2\sum_{\vec r, \vec r\,^{\prime}}
m(\vec r) G^{(2)}(\vec r -\vec r\,^{\prime}) m(\vec r\,^{\prime})
\Biggr], \eqno (4)$$
where the pre-factor of $\beta^{N_m}$ above results from counting all
possible instanton configurations fixed at 
$N_m$ sites in space.
Hence, we obtain a transition at 
$k_B T_c \gsim k_B T_0 \cong 0.44 g_0 \hbar\omega_0$ that is dual to the
2D CG transition,$^{23}$
where $T_0$ denotes the temperature at which the chemical
potential vanishes in the CG ensemble (4).
For temperatures below $T_c$ free instantons  exist,
while instanton-anti-instanton pairs are bound above $T_c$.
Note that the exponentially small number of free instantons
that exist near  the dual CG 
transition$^{23}$ validates the previous
assumption of diluteness.

To demonstrate that the preceding is in fact a CD-transition, 
we now compute
the confinement length-scale (string-tension), and show
that it diverges (vanishes)
exponentially as $T_c$ is approached from low-temperature.
Adapting Polyakov's zero-temperature
calculation$^{20,21}$ for the auto-correlation function
of the electromagnetic fields in  compact QED$_3$
to the present case at $T\neq 0$ reduces to the sine-Gordon 
reformulation of the dual
2D CG ensemble (4).$^{23}$  A straight-forward generalization
of his methods then yields that the corresponding
static  auto-correlation
function for the statistical electrodynamic field is given by
$g_0^{-2}\beta\langle b_i(\vec k) b_j(-\vec k)\rangle =
\delta_{ij} - k_i  k_j/[k^2+\xi_{\chi}^{-2}(T)]$, where
the magnetic fields in the spatial directions $j = 1,2$ are related
to the electric field $e_l(x)=F_{0,l}(x)$ by
$b_j(x)=i\epsilon_{jl}e_l(x)$, and where
$\xi_{\chi}(T)$ denotes the Debye
screening-length of the dual CG ensemble (4).
Hence, the confinement length scale $\xi_{\chi}(T)$ 
is infinite  above $T_c$ in the bound-instanton phase,
while   it diverges below $T_c$
like 
$$\xi_{\chi}(T) = a B\, {\rm exp}[A/(1-T/T_c)^{1/2}]\eqno (5)$$
 in the free instanton
phase, with $A$ and $B$ being non-universal 
numerical constants.$^{23}$  A corresponding
adaptation of Polyakov's calculation of the
zero-temperature string-tension$^{20,21}$ $\sigma$ 
also yields
$$\sigma(T)\sim g_0^2 \hbar\omega_0 \xi_{\chi}^{-1}(T).\eqno (6)$$
In
conclusion, correlations between the statistical gauge
field have a finite range below $T_c$ indicating confinement,
while they are of infinite-range above $T_c$ implying gaussian
[$n_{\mu\nu}(x)=0$] electrodynamics consistent 
with the absence of instantons.  
This observation suggests
that the specific-heat contribution of the gauge-fields in 
the weak-coupling limit of
compact QED$_3$ follows a gaussian (photon) $T^2$ law above $T_c$, 
while  the temperature-scale
below which activated behavior sets in the
confined phase
is on the order of  $\sigma (0) a$.
Here, the zero-temperature string tension $\sigma (0)$
is given by (6)  with a zero-temperature confinement 
length-scale$^{20,21}$  of
$\xi_{\chi}(0) = 
a (g_0/2\pi)  e^{{\rm const}/g_0^2}$.  The   
former energy-scale is exponentially small in the present
weak-coupling limit, however, indicating that the specific
heat is given essentially  by the  same gaussian $T^2$ law
below $T_c$.

{\it Strong-coupling.}  Following Polyakov and Susskind,$^{14,15,21}$
the Hamiltonian formulation of compact QED$_3$ (1)
reduces to the kinetic energy 
$H_0={1\over 2}g_0^{-2}\hbar\omega_0\sum_{\vec x}e_i^2(\vec x)$,
in the regime $g_0\gg 1$, where the statistical electric field
operator $e_i(\vec x)=-ig_0^2\partial/\partial a_i(\vec x)$
satisfies Maxwell's  Eq., $\Delta_i e_i = 0$, in vacuum. (The temporal
gauge $a_3 = 0$ has implicitly been chosen.)   Because of the 
compact nature of the vector potential, 
the field-strength operator $g_0^{-2} e_i(\vec x)$
has integer eigenvalues $n_i(\vec x)$ satisfying 
$\Delta_i n_i(\vec x)=0$.  Since the latter constraint can be
easily solved by taking $n_i(\vec x)=\epsilon_{ij}\Delta_j n(\vec r)$,
where $n(\vec r)$ is an integer-field on the dual square-lattice,
we arrive at the discrete gaussian (DG) model for surface
roughening$^{24-26}$ given by
$$Z=\sum_{\{n(\vec r)\}} {\rm exp}
\Biggl\{-{1\over 2}\beta g_0^2\sum_{\vec r}
[\Delta_i n(\vec r)]^2\Biggr\}.\eqno (7)$$
This model is dual to the 2D CG ensemble,$^{24}$ with
a transition temperature  
$k_B T_c \cong 0.73 g_0^2\hbar\omega_0$.
Notice also that the present DG model (7) is equal to the
weak-coupling instanton action (2) for the special
case of static configurations
$\Delta_3 n(\bar x)=0$.  Below we demonstrate
that $T_c$ marks the boundary
for the CD transition  that also
exists
in the strong-coupling limit.$^{16,17}$  

Let us  first determine the
temperature dependence of the
confinement length-scale for compact
QED$_3$ in the present strong-coupling limit.
Given the identity 
$g_0^{-2}\langle e_k(\vec x) e_l(\vec x^{\prime})\rangle
= g_0^2\epsilon_{ki}\epsilon_{lj}
\langle \Delta_i n(\vec r) \Delta_j n(\vec r\,^{\prime})\rangle$,
and the fact that the latter  is ``proportional''
to the height-height 
correlation function of the surface-roughening model (7),$^{24}$
we find that the spatial Fourier transform of the
auto-correlation
function for the statistical electric-field is given by
$g_0^{-2}\langle e_k(\vec k) e_l(-\vec k)\rangle
= \beta^{-1}\epsilon_{ki}\epsilon_{lj}
k_i k_j/[k^2+\xi_{\chi}^{-2}(T)]$,
where $\xi_{\chi}$ denotes
the Debye screening length
 of the
dual CG ensemble.  Hence,
we recover precisely the same pole obtained in the
prior weak-coupling analysis, with a confinement length-scale 
$\xi_{\chi}(T)$ that 
diverges exponentially (5) at $T_c$,
but that vanishes linearly with
temperature
at
zero temperature.$^{23}$
Note that $g_0^{-2}\beta\langle\vec e(0)\cdot\vec e(0)\rangle$
jumps from zero to unity upon heating through the
transition.

Confinement properties
are also directly probed by the string-tension.
To compute  this quantity, 
it is convenient to return to the original
strong-coupling Hamiltonian that is known to be 
dual to the 2D $XY$-model.
Specifically, Polyakov and Susskind have shown that the
effective potential energy between opposing unit statistical
charges separated by a distance $R$ is given by
$V(R)=-k_B T\,{\rm ln}\langle{\rm exp}
\, i[\phi(0)-\phi(\vec R)]\rangle$, 
where $\phi(\vec R)$ denotes the phase variable of the dual
XY-model.$^{14,15}$  Yet since the correlation function for this
model is given by
$\langle{\rm exp}\, i[\phi(0)-\phi(\vec R)]\rangle
=(r_0/R)^{1/4}{\rm exp}(-R/\xi_{\chi})$  near $T_c$, 
we have that the string-tension
[defined by ${\rm lim}_{R\rightarrow\infty} V(R)=\sigma R$] is
$\sigma(T) = k_B T_c  \xi_{\chi}^{-1}(T)$
in this region.  
[This result agrees with  Eq. (6).] 
On the other hand,
the zero-temperature string-tension in the strong-coupling
limit is given by 
$\sigma(0)={1\over 2} g_0^2 \hbar\omega_0 a^{-1}$,
since a confining flux-tube corresponds to a smooth
step in the DG-model for surface-roughening (7).
In fact, it is evident that the step free-energy per length
of this model is equivalent to the string-tension 
in general.  The temperature-dependence  of the former quantity 
is well known,$^{25,26}$ and  agrees with the results  just
cited.  

Last, the specific-heat per site for the 
interface roughening
model (7) obtained from Monte-Carlo simulations$^{25}$ is
reproduced in Fig. 1.  It is qualitatively
similar to the weak-coupling result
discussed previously, with the exception
that ({\it i}) the strong-coupling 
specific-heat saturates to the classical value of
${1\over 2}k_B$ above the transition 
because  $k_B T_c\gg\hbar\omega_0$, and that ({\it ii})
the energy-scale for activated behavior is on
the order of $\sigma(0) a \cong 0.69 k_B T_c$.  The latter
observation  confirms the fact that the elementary excitations
in  compact QED$_3$ at low-temperature are strings of electric flux.

{\it Anyon Superconductors and Spin-liquids.}  We now apply
these results  to the spin-1/2 anyon 
superconductor,$^{11}$ which is
known to be a large-$N$ saddle-point  of the
2D $t-J$ model near half-filling.$^{12}$  
Consider then ideal spin-1/2 psuedo-fermions 
on the square-lattice with a statistical flux-tube of
one flux-quantum attached to each.  The corresponding
long-wavelength limit CS 
lagrangian is given by
$$\eqalignno{L=\sum_i \Biggl\{
{i\over 2}(\bar f_{i\sigma}\dot f_{i\sigma} -{\rm c.c.})
-&(\tilde t e^{i [A_0+a]_{\mu}(i)}
\bar f_{i\sigma} f_{i+\hat\mu,\sigma}+{\rm c.c.}) +
(2\pi)^{-1}\epsilon_{\mu\nu} a_{\mu}(i)\dot a_{\nu}(i)
+\cr
&+(8\pi)^{-1}\epsilon_0 \dot a_{\mu}^2(i) -
(8\pi\mu_0)^{-1} [\Delta_1 a_2(i) - \Delta_2 a_1(i)]^2\Biggr\} 
& (8) \cr}$$
in the temporal gauge $a_3(i)=0$, where $\tilde t$ denotes the
hopping matrix-element for the psuedo-fermion ($f_{i\sigma}$) and
$\dot a_{\nu}(i)=\hbar\partial  a_{\nu}(i)/\partial t$,
etc.. 
Also,
$\epsilon_0$ and $\mu_0$ respectively  denote the vacuum dielectric
constant and the magnetic permiability for fluctuations
$a_{\mu}(i)$ of the statistical gauge-field,$^{11,12}$
while
$\vec A_0$  describes the statistical magnetic field
due to the   flux-tubes averaged over the entire
lattice.$^{9,10}$  After integrating out the psuedo-fermions,
one finds that
the cancellation of the resulting  Hall conductance against that
of the CS term leads to the action (1) for compact
QED$_3$ at zero-temperature,$^{12,27}$ 
with a zero-sound velocity  
$c_0=\hbar^{-1}(\mu_0^{-1}+\mu_f^{-1})^{1/2}/(\epsilon_0+\epsilon_f)^{1/2}$
and a coupling-constant
$g_0^{-2}=(2\pi)^{-1}(\epsilon_0+\epsilon_f)\hbar\omega_0$.  Here,
$\epsilon_f\sim \Delta_f^{-1}$ and $\mu_f^{-1}\sim\Delta_f\xi_{f}^2$
are the psuedo-fermion contribution
to the dielectric constant and the magnetic permiability, 
respectively, with $\Delta_f$ and
$\xi_{f}$ denoting the 
relevant (Hofstadter) energy-gap and magnetic length of 
the psuedo-fermions
in the
mean-field approximation.$^{11}$  Now suppose that 
we extend this result
above zero-temperature.  Then 
no {\it free} statistical charge can exists below 
$T_c$ because of confinement,
implying that the Hall conductance generated by the psuedo-fermions
is temperature-independent.  Due to the exact cancellation of
the latter against the CS-term,  we ({\it i}) 
recover the assumption that
the statistical gauge-field is described by 
vacuum QED$_3$ below $T_c$,
and find ({\it ii}) that the electromagnetic 
response is that of a superconductor in this 
regime.$^{11,12}$  Hence, $T_c$ is in fact the critical
temperature for lattice anyon-superconductors.  
This is in sharp
contrast to  straight-forward extensions of
RPA methods  to non-zero temperature,$^{28}$
which  find that the critical temperature 
vanishes due to a spurious temperature dependence
acquired by the psuedo-fermion Hall conductance.  Zero-temperature  
RPA calculations find,$^{11,12}$ however, that the
London penetration length  varies as
$\lambda_L\propto (\epsilon_0+\epsilon_f)^{1/2}$.  
By extension,
$\lambda_L^{-1}$ should have an  $s$-wave-like 
temperature dependence
at low-temperature,
while it should  {\it jump} to zero at the transition.
In conclusion,  particle-hole excitations 
in lattice anyon superconductors
are bound by a confining string.$^{4,12}$ 
The similarity between the duality
of this transition  to that of the 2D $XY$
model$^{14,15}$  
and the duality of conventional 
superconductors in the presence of fluctuating
magnetic fields to the three-dimensional $XY$ model$^{29}$ is 
striking.
Also, if we assume that 
$k_B T_c\ll\Delta_f$,  
then the specific-heat should
be dominated by the gauge-field contribution shown in Fig. 1.
Such ``rounding'' 
of the specific-heat anomaly  actually occurs
in certain high-temperature superconductors.$^{18}$

In the case of the quantum-disordered (spin-liquid) phase 
of the 2D quantum antiferromagnet, on the other hand,
$CP_1$ calculations indicate that chiral spin-fluctuations$^8$
are also described by the action (1)
for vacuum QED$_3$, where 
$c_0$ is identified here with the spin-wave velocity.$^{6,7}$
The  coupling-constant is given by
$g_0^2=12\pi \Delta_f/\hbar\omega_0$ in such case, 
where $\Delta_f$ denotes
the spin-gap.$^7$   Hence, spinons are
confined below $T_c$, where the elementary
excitations are again strings of statistical electric
flux.  Also, the
existence of a universal-jump 
in the long-wavelength autocorrelation for the electric fields
implies a corresponding jump 
$\Delta\langle Q_{\chi}^2\rangle/{\cal N}
=(2\pi)^{-2}g_0^2/\beta_c$
in the fluctuation per (${\cal N}$) site
 of the total number 
of skyrmions  (quanta of chiral spin) 
$Q_{\chi} =  (2\pi)^{-1}\sum_{\vec x}
\epsilon_{ij} \Delta_i a_j (\vec x)$
at the transition.$^{7,21}$  
Last, the gauge-fields will contribute
a smooth bump anomaly to the specific
heat below $T_c$ (see Fig. 1), 
just as in the previous case.

To  conclude, it has been demonstrated
that the CD transition shown by compact QED$_3$ is dual
to the 2D CG transition in both the weak and the strong
coupling limits.  It has also been argued that the corresponding
transition temperature coincides with the critical temperature
of  ideal  anyon superconductors and of spin-liquids
on the square-lattice.
The former constitutes a first step towards the thermodynamics of
anyon superconductors, which are perhaps the paradigm
for 2D superconductivity in general.

The author is grateful to R. Laughlin, E. Rezayi, 
A. Sokol, S. Trugman, J. Engelbrecht 
and S. Hebboul for stimulating discussions.    This work
was performed under the auspices of the U.S. Department of Energy
and  was supported in part by National
Science Foundation grant DMR-9322427.

\vfill\eject
\centerline{\bf References}
\vskip 16 pt

\item {1.} {\it The Physical Properties
of High-Temperature Superconductors},
vol. 2, edited
by D.M. Ginsberg (World Scientific, Singapore, 1990).

\item {2.}  P.W. Anderson, Science
 {\bf 235}, 1196 (1987).

\item {3.} R.B. Laughlin, Phy. Rev. Lett. {\bf 60}, 2677 (1988).

\item {4.} R.B. Laughlin, Science {\bf 242}, 525 (1988).

\item {5.} D.P. Arovas and A. Aurbach, Phys. Rev. B{\bf 38}, 316
(1988).

\item {6.} N. Read and S. Sachdev, Phys. Rev. Lett. {\bf 62}, 1694
(1989).

\item {7.} J.P. Rodriguez, Phys. Rev. B {\bf 41},
7326 (1990).

\item {8.} X.G. Wen, F. Wilczek, and A. Zee, Phys. Rev. B {\bf 39},
11413 (1989).

\item {9.} A. Fetter, C. Hanna,
R.B. Laughlin, Phys. Rev. B {\bf 39},
9679 (1989).

\item {10.} Y. Chen, F. Wilczek,
E. Witten, and B.I. Halperin, Int. J. 
Mod. Phys. B {\bf 3}, 1001 (1989).

\item {11.} P.B. Wiegmann, Phys. Rev. Lett. {\bf 65}, 2070 (1990);
J.P. Rodriguez. and B. Dou\c cot,
Phys. Rev. B{\bf 42}, 8724 (1990); (E) {\bf 43}, 6209 (1991).

\item {12.}  J.P. Rodriguez and B. Dou\c cot,
Phys. Rev. B{\bf 45}, 971 (1992).

\item {13.} L.B. Ioffe and A.I. Larkin,
 Phys. Rev. B {\bf 39}, 8988 (1989).

\item {14.} A. Polyakov, Phys. Lett.  {\bf 72B}, 477 (1978).

\item {15.} L. Susskind, Phys. Rev. D {\bf 20}, 2610 (1979).

\item {16.} I. Ichinose and T. Matsui, Nucl. Phys. 
{\bf B394}, 281 (1993).

\item {17.} N. Nagaosa, Phys. Rev. Lett. {\bf 71}, 
4210 (1993).

\item {18.} J.W. Loram et al., Phys. Rev. Lett. {\bf 71}, 
1740 (1993);
H. Wuhl et al., Physica C {\bf 185 - 189}, 755 (1991); 
A. Junod in ref. 1.

\item {19.}  A.A. Abrikosov, L.P. Gorkov, 
and I.E. Dzyaloshinski, {\it Methods of 
Quantum Field Theory in Statistical Physics}, 
(Dover, New York, 1975).

\item {20.} A.M. Polyakov, Nucl. Phys. B{\bf 120}, 429 (1977).

\item {21.} A.M. Polyakov, {\it Gauge Fields and Strings} 
(Harwood, New York, 1987).

\item {22.} T. Banks, R. Myerson and J. Kogut, Nucl. Phys. {\bf B129},
493 (1977).

\item {23.} P. Minnhagen, Rev. Mod. Phys. {\bf 59}, 1001 (1987).

\item {24.} S.T. Chui and J.D. Weeks, Phys. Rev. 
B {\bf 14}, 4978 (1976);
J.D. Weeks, in 
{\it Ordering in Strongly Fluctuating Condensed Matter Systems},
edited by T. Riste (Plenum, New York, 1980).

\item {25.} R.H. Swendsen, Phys. Rev. B {\bf 15}, 5421 (1977).

\item {26.} H. van Beijeren and I. Nolden, in
{\it Structure and Dynamics of Surfaces II}, 
edited by W. Schommers
and P. von Blanckenhagen (Springer, Heidelberg, 1987).

\item {27.}  It is shown in ref. 12 that the effective action
for the statistical gauge-field is 
time-reversal/parity symmetric to quadratic order in frequency
and wave-number.  This contrasts with the 
physical electromagnetic response, which 
exhibits a zero-field Hall
effect within RPA.

\item {28.} J.E. Hetrick, Y. Hosotani, and B.H. Lee, 
Ann. Phys. (NY) {\bf 209}, 151 (1991).

\item {29.} C. Dasgupta and B.I. Halperin, Phys. Rev. Lett.
{\bf 47}, 1556 (1981).

\vfill\eject
\centerline{\bf Figure Caption}
\vskip 20pt
\item {Fig. 1.}   Shown are Monte-Carlo results 
obtained 
by Swendsen (ref. 25) for
the specific heat of the DG model (7) 
on a $10\times 10$ lattice.  The specific-heat is
plotted in units of $k_B$, while the temperature
is plotted in units of $g_0^2\hbar\omega_0/k_B$.  The
circles represent extrapolations based on the
solid-on-solid model (ref. 25). 

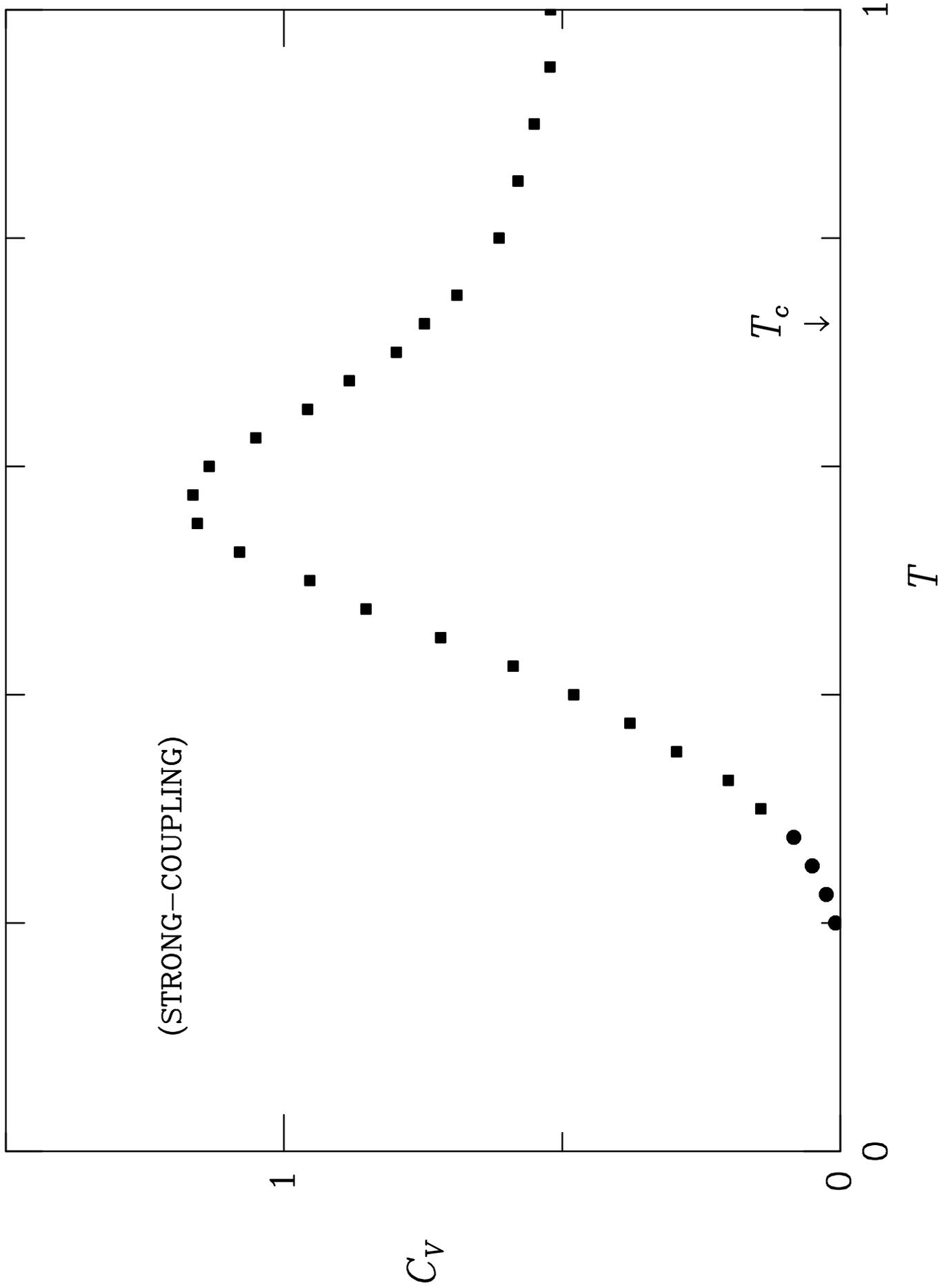
\end